\documentclass[english,aps,citeautoscript,elide,prb,showpacs,twocolumn]{revtex4-1}
\usepackage{mathptmx}
\usepackage[T1]{fontenc}
\usepackage[latin9]{inputenc}
\usepackage{color}
\usepackage{babel}
\usepackage{prettyref}
\usepackage{amsmath}
\usepackage{amssymb}
\usepackage{graphicx}
\usepackage{esint}
\usepackage[unicode=true,
 bookmarks=true,bookmarksnumbered=false,bookmarksopen=true,bookmarksopenlevel=3,
 breaklinks=true,pdfborder={0 0 0},backref=false,colorlinks=true]
 {hyperref}
\hypersetup{pdftitle={Photoinduced pure spin current injection in graphene with Rashba spin-orbit interaction},
 pdfauthor={Julien Rioux},
 pdfsubject={graphene; Rashba interaction; spin current injection; spin-orbit coupling},
 citecolor=blue,linkcolor=blue,pdfpagemode=UseNone,pdfstartview=FitH,urlcolor=blue}

\makeatletter

 \@ifundefined{textcolor}{}
 {%
   \definecolor{BLACK}{gray}{0}
   \definecolor{WHITE}{gray}{1}
   \definecolor{RED}{rgb}{1,0,0}
   \definecolor{GREEN}{rgb}{0,1,0}
   \definecolor{BLUE}{rgb}{0,0,1}
   \definecolor{CYAN}{cmyk}{1,0,0,0}
   \definecolor{MAGENTA}{cmyk}{0,1,0,0}
   \definecolor{YELLOW}{cmyk}{0,0,1,0}
 }

\usepackage{afterpage}
\usepackage{hypcap}
\usepackage{svn}
\usepackage{times}
\usepackage{upgreek}

\SVN $Date: 2014-07-28 14:58:55 -0400 (Mon, 28 Jul 2014) $
\SVN $Revision: 706 $

\renewcommand{\vec}[1]{\boldsymbol{\mathbf{#1}}}

\AtBeginDocument{\renewcommand{\eqref}[1]{\hyperref[#1]{(\ref*{#1})}}}
\newrefformat{eq}{\hyperref[#1]{Eq.~(\ref*{#1})}}
\newrefformat{eqs}{Eqs.~\hyperref[#1]{(\ref*{#1})}}
\newrefformat{equation}{\hyperref[#1]{Equation~(\ref*{#1})}}
\newrefformat{fig}{\hyperref[#1]{Fig.~\ref*{#1}}}
\newrefformat{figure}{\hyperref[#1]{Figure~\ref*{#1}}}
\newrefformat{secapp}{\hyperref[#1]{Appendix}}
\newrefformat{sec}{\hyperref[#1]{Sec.~\ref*{#1}}}
\newrefformat{subfiga}{\hyperref[#1]{Fig.~\ref*{#1}(a)}}
\newrefformat{subfigb}{\hyperref[#1]{Fig.~\ref*{#1}(b)}}
\newrefformat{subfigc}{\hyperref[#1]{Fig.~\ref*{#1}(c)}}
\newrefformat{subfigd}{\hyperref[#1]{Fig.~\ref*{#1}(d)}}
\newrefformat{subfigab}{\hyperref[#1]{Fig.~\ref*{#1}(a,b)}}
\newrefformat{subfigcd}{\hyperref[#1]{Fig.~\ref*{#1}(c,d)}}
\newrefformat{tab}{\hyperref[#1]{Table~\ref*{#1}}}

\allowdisplaybreaks

\makeatother

\begin{document}

\preprint{Revision \SVNRevision}

\title{Photoinduced pure spin current injection in graphene with Rashba
spin-orbit interaction}

\author{Julien Rioux}

\affiliation{Department of Physics, University of Konstanz, D-78457 Konstanz,
Germany}

\author{Guido Burkard}

\affiliation{Department of Physics, University of Konstanz, D-78457 Konstanz,
Germany}

\date{\SVNDate}
\begin{abstract}
We propose a photoexcitation scheme for pure spin current generation
in graphene subject to a Rashba spin-orbit coupling. Although excitation
using circularly-polarized light does not result in optical orientation
of spins in graphene unless an additional magnetic field is present,
we show that excitation with linearly-polarized light at normal incidence
yields spin current injection without magnetic field. Spins are polarized
within the graphene plane and are displaced in opposite directions,
with no net charge displacement. The direction of the spin current
is determined by the linear polarization axis of the light, and the
injection rate is proportional to the intensity. The technique is
tunable via an applied bias voltage and is accessible over a wide
frequency range. We predict a spin current polarization as high as
$75$\% for photon frequencies comparable to the Rashba frequency.
Spin current injection via optical methods removes the need for ferromagnetic
contacts, which have been identified as a possible source of spin
scattering in electrical spin injection in graphene.
\end{abstract}



\maketitle

\global\long\def\bra#1{\left\langle #1\right|}
\global\long\def\ket#1{\left|#1\right\rangle }
\global\long\def\tensor#1{\vec{#1}}

\global\long\def\cc{\mathrm{c.c.}}
\global\long\def\d{\mathrm{d}}
\global\long\def\Hamiltonian{H}
\global\long\def\Heaviside{\Theta}
\global\long\def\phase{\varphi}
\global\long\def\disparity{d}

\setcounter{topnumber}{0}
\afterpage{\setcounter{topnumber}{1}}

\section{Introduction}

The isolation of graphene, a single layer of graphite, has opened
the door to research on atomically thin crystals with Dirac-like electrons
\citep{Novoselov2004}. In addition to the high mobility of charge
carriers in graphene, rendering it attractive for use in electronics,
long spin relaxation times are expected due to the spinless atomic
nucleus of $^{12}$C and the small spin-orbit coupling, further promoting
graphene as an interesting material for spintronics applications \citep{Trauzettel2007b,Abergel2010}.

The first studies of spin injection in graphene have reported a diffusive
spin current injected via ferromagnetic contacts, initially in a two-terminal
geometry \citep{Hill2006} and soon after in a nonlocal four-terminal
geometry \citep{Ohishi2007,Tombros2007,Cho2007}. The technique is
improved when inserting MgO as an insulating tunnel barrier \citep{Han2010}
and achieves up to 60\% spin polarization when using a second layer
of graphene as the tunnel barrier \citep{Friedman2014}. Thus far,
the obtained spin lifetimes are low compared to expectations, and
the ferromagnetic contacts remain a possible source of spin scattering
explaining this discrepancy \citep{Volmer2013}. The efforts towards
electrical spin injection in graphene are reviewed by Shiraishi \citep{Shiraishi2014}.
Others have reported spin current injection in graphene using dynamical
\citep{Patra2012,Tang2013} and thermal \citep{Zeng2011c,Torres2014}
methods.

Optical orientation is another means of injecting spin-polarized carriers
into semiconductors exhibiting spin-orbit coupling \citep{Dyakonov1984}.
Spin photocurrents resulting from absorption of linearly-polarized
light have been proposed and demonstrated in GaAs quantum wells \citep{Tarasenko2005a,Sherman2005a,Bhat2005b,Zhao2005}.
Although graphene spintronics offers very interesting prospects, the
optical injection and control of spin currents in graphene have yet
to be investigated. Optical methods are motivated by the relatively
strong 2.3\% absorption of light by a single graphene sheet over a
wide range of frequencies \citep{Nair2008}. Since light interacts
with the orbital degree of freedom, any optical manipulation of the
spin degree of freedom relies on the presence of coupling between
these two degrees of freedom. Graphene's weak spin-orbit interaction
(SOI), which limits the application range for spin photoinjection
and, conversely, for spin readout by optical methods \citep{Pesin2012},
is increased through heavy-atom intercalation \citep{Marchenko2012},
impurities \citep{CastroNeto2009b}, or hydrogenation \citep{Balakrishnan2013,Gmitra2013}.
In contrast to intrinsic SOI, Rashba-type SOI is tunable and can be
increased to a strength sufficiently large that optical spin injection
in graphene with an in-plane magnetic field has recently been proposed
\citep{Inglot2014}.

In this paper, optical spin current injection is investigated in graphene
subject to a Rashba spin-orbit coupling. Due to the chirality of graphene
electrons and the nature of the optical matrix element, we show that
photoexcitation results in the injection of a pure spin current, without
accompanying charge current. We thus predict spin current injection
in graphene via optical methods without contacts, eliminating a possible
source of spin scattering. The paper is organized as follows. The
effective Hamiltonian and matrix elements used for the calculations
are presented in \prettyref{sec:Hamiltonian}. Spin current injection
and its polarization are calculated in \prettyref{sec:Results}. We
summarize and discuss our results in \prettyref{sec:Summary}.

\section{Hamiltonian and matrix elements \label{sec:Hamiltonian}}

\begin{figure}
\capstart%

\includegraphics{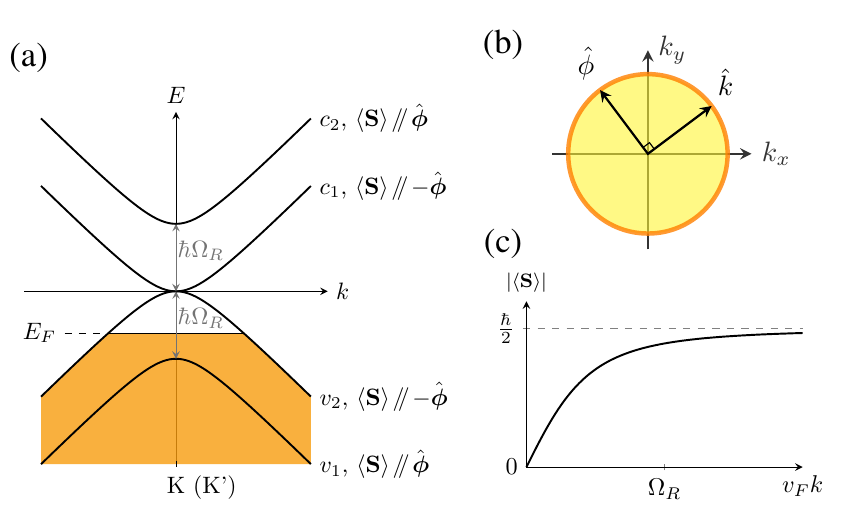}

\caption{(a) Energy-momentum dispersion of graphene with Rashba spin-orbit
coupling; (b) vectors defining the orientation in the graphene plane;
and (c) magnitude of the spin expectation value as a function of crystal
momentum. \label{subfiga:dispersion} \label{subfigc:spin-expectation-value}
\label{subfigb:top-view}}
\end{figure}

Band electrons in single-layer graphene are described by the usual
Dirac Hamiltonian in the linear dispersion regime \citep{Kane2005c},
\begin{equation}
\Hamiltonian_{0}=\hbar v_{F}\left(\tau\sigma_{x}k_{x}+\sigma_{y}k_{y}\right),
\end{equation}
where $v_{F}$ is the Fermi velocity, $\vec{\sigma}$ are the Pauli
matrices, here acting on graphene's A and B sublattice space, and
$\vec{k}$ is the in-plane crystal momentum relative to the K~point
($\tau=1$) or the K'~point ($\tau=-1$). An external out-of-plane
electric field breaks inversion symmetry and introduces the Rashba
spin-orbit coupling term
\begin{equation}
\Hamiltonian_{R}=\hbar\Omega_{R}\left(\tau\sigma_{x}S_{y}-\sigma_{y}S_{x}\right),
\end{equation}
where $\Omega_{R}$ is the Rashba frequency and $\vec{S}$ is the
electron spin \citep{Kane2005c}. This can be induced by the substrate
or by additional gates generating a voltage gradient perpendicular
to the sample. The result of diagonalizing $\Hamiltonian_{0}+\Hamiltonian_{R}$
yields four energy bands with an isotropic band dispersion quadratic
in $k$ for $k\ll\Omega_{R}/2v_{F}$ and linear in $k$ for $k\gg\Omega_{R}/v_{F}$,
shown in \prettyref{subfiga:dispersion}. The lowest-energy conduction
band $c_{1}$ and highest-energy valence band $v_{2}$ are degenerate
at the K~point, and so-called ``split-off'' conduction band $c_{2}$
and valence band $v_{1}$ are respectively shifted up and down by
an energy $\hbar\Omega_{R}$ from the charge neutrality point \citep{Rashba2009a}.
In the two gapless bands the expectation value of the electron spin,
$\left\langle \vec{S}\right\rangle $, is oriented antiparallel with
$\hat{\phi}=\hat{z}\times\hat{k}$, where $\hat{k}$ is the unit vector
parallel to the direction of $\vec{k}$ {[}\emph{cf.}~\prettyref{subfigb:top-view}{]}
and $\hat{z}$ is normal to the graphene plane, while $\left\langle \vec{S}\right\rangle $
is oriented parallel with $\hat{\phi}$ for the split-off bands. In
all cases, the expectation value of spin has a magnitude
\begin{equation}
\left|\left\langle \vec{S}\right\rangle \right|=\frac{v_{F}\hbar k}{\sqrt{\Omega_{R}^{2}+4v_{F}^{2}k^{2}}},\label{eq:spin-expectation-value}
\end{equation}
shown in \prettyref{subfigc:spin-expectation-value}, reaching $\hbar/2$
for large $k\gg\Omega_{R}/v_{F}$.

The interaction of the material with an external vector potential
$\vec{A}$ describing an electromagnetic field is treated within minimal
coupling by substituting $\hbar\vec{k}$ by $\hbar\vec{k}-e\vec{A}$
in the Hamiltonian. The linear response under photoexcitation is readily
obtained from the matrix elements, between initial and final states,
of the $\vec{A}\cdot\vec{v}$ interaction term appearing in minimal
coupling. Matrix elements of the velocity operator $\vec{v}=\vec{\sigma}v_{F}$
and of the spin operator are given in the \prettyref{secapp:Matrix-elements}.

\section{Spin current injection in graphene \label{sec:Results}}

In this section, we consider photoexcitation of a graphene layer with
Rashba SOI under a monochromatic electric field
\begin{equation}
\vec{E}(t)=\vec{E}(\omega)e^{-i\omega t}+\vec{E}^{*}(\omega)e^{i\omega t}\,.
\end{equation}
The $\vec{k}$-dependent spin polarization of the eigenstates {[}\emph{cf.}~\prettyref{eq:spin-expectation-value}{]}
is exploited to inject photoinduced spin currents. The spin textures
in the K and K' valleys are exactly the same, thus carriers of momentum
$\vec{k}$ near K and K' are excited with equal spin polarization.
Moreover, carriers at $-\vec{k}$ are excited with opposite spin polarization,
resulting in a pure spin current at normal incidence.

We use the generally accepted symmetrized spin current operator $\tensor J_{s}=\frac{1}{2}\left(\vec{v}\vec{S}+\vec{S}\vec{v}\right)$
\citep{Rashba2004b}, where juxtaposed vectors form the dyadic product.
The injection rate $\dot{\tensor J}_{s}$ is derived by solving the
Heisenberg equation of motion and keeping the nonzero term at lowest
order in the field. The resulting spin current injection rate is linear
in intensity and is written in component notation \citep{Bhat2005b,note-on-units-system}
\begin{equation}
\dot{J}_{s}^{ab}=\mu_{1}^{abcd}(\omega)E^{c*}(\omega)E^{d}(\omega),\label{eq:mu1-definition}
\end{equation}
where Roman superscripts indicate Cartesian components and repeated
superscripts are summed over. The response pseudotensor $\mu_{1}^{abcd}(\omega)$,
derived at a level equivalent to Fermi's golden rule and including
both contributions of electrons and holes, is
\begin{multline}
\mu_{1}^{abcd}(\omega)=\frac{2\pi e^{2}}{\hbar^{2}\omega^{2}}\sum_{cv}\int\frac{\d^{2}k}{4\pi^{2}}\left[\left(J_{s}\right)_{cc}^{ab}-\left(J_{s}\right)_{vv}^{ab}\right]\\
\times v_{cv}^{c*}(\vec{k})v_{cv}^{d}(\vec{k})\,\delta[\omega_{cv}(\vec{k})-\omega],\label{eq:mu1-microscopic}
\end{multline}
where
\begin{equation}
\left(J_{s}\right)_{cc}^{ab}(\vec{k})\equiv\sum_{m}\tfrac{1}{2}[v_{cm}^{a}(\vec{k})S_{mc}^{b}(\vec{k})+S_{cm}^{a}(\vec{k})v_{mc}^{b}(\vec{k})]\label{eq:J-matrix-element}
\end{equation}
is the diagonal matrix element of the spin current operator, $\vec{v}_{mn}(\vec{k})$
and $\vec{S}_{mn}(\vec{k})$ indicate matrix elements of, respectively,
the velocity and spin operators, and $\omega_{cv}(\vec{k})\equiv[E_{c}(\vec{k})-E_{v}(\vec{k})]/\hbar$
is the energy difference between conduction and valence bands; the
sum over $m$ includes all band indices, while the sum over $c$ ($v$)
includes conduction (valence) bands only. Within the Dirac model of
\prettyref{sec:Hamiltonian}, only one independent nonzero component
exists: $\mu_{1}^{xyxx}(\omega)$. There are in total eight nonzero
components, related by 
\begin{multline}
\mu_{1}^{xyxx}=\mu_{1}^{yxxx}=\mu_{1}^{yyxy}=\mu_{1}^{yyyx}\\
=-\mu_{1}^{xxxy}=-\mu_{1}^{xxyx}=-\mu_{1}^{xyyy}=-\mu_{1}^{yxyy}.
\end{multline}
Evaluating the charge current injection in the same approach yields
a vanishing response tensor $\eta_{1}$ \citep{Bhat2005b}, thus the
spin current presented here is pure, without accompanying charge current.

A general oscillatory electric field for a normally-incident wave
has the form $\vec{E}(\omega)=E_{\omega}e^{i\varphi_{\omega}}\left(\hat{\vec{x}}_{\omega}+\hat{\vec{y}}_{\omega}e^{i\delta\varphi_{\omega}}\right)/\sqrt{2}$,
where the amplitude $E_{\omega}$ and the phase parameters $\varphi_{\omega}$
and $\delta\varphi_{\omega}$ are chosen to be real for an appropriate
choice of orthonormal vectors $\hat{\vec{x}}_{\omega}$ and $\hat{\vec{y}}_{\omega}$
in the graphene plane.  From \prettyref{eq:mu1-definition} and the
symmetry of $\mu_{1}^{abcd}(\omega)$ given above, the injection rate
of the spin current is given by the general dyadic expression
\begin{equation}
\dot{\tensor J}_{s}=\mu_{1}^{xyxx}(\omega)\left|E_{\omega}\right|^{2}\cos(\delta\varphi_{\omega})\left(\hat{\vec{y}}_{\omega}\hat{\vec{y}}_{\omega}-\hat{\vec{x}}_{\omega}\hat{\vec{x}}_{\omega}\right).\label{eq:Jdot-general}
\end{equation}
Due to the cosine dependence on the Stokes parameter $\delta\varphi_{\omega}$,
the spin current injection is zero for circularly-polarized light
($\delta\varphi_{\omega}=\pm\frac{\pi}{2}$) and maximum for linearly-polarized
light ($\delta\varphi_{\omega}=0$), in which latter case we obtain
\begin{equation}
\dot{\tensor J}_{s}=\mu_{1}^{xyxx}(\omega)\left|E_{\omega}\right|^{2}\left(\hat{\vec{e}}_{\omega}\hat{\vec{e}}_{\omega}^{\perp}+\hat{\vec{e}}_{\omega}^{\perp}\hat{\vec{e}}_{\omega}\right),\label{eq:Jdot-linear}
\end{equation}
where $\hat{\vec{e}}_{\omega}\equiv\left(\hat{\vec{x}}_{\omega}+\hat{\vec{y}}_{\omega}\right)/\sqrt{2}$
is the linear polarization axis and $\hat{\vec{e}}_{\omega}^{\perp}=\hat{\vec{z}}\times\hat{\vec{e}}_{\omega}$
is an in-plane unit vector perpendicular to the polarization axis.
Equation \eqref{eq:Jdot-linear} is the main result of this paper.
Linearly-polarized light induces a spin current injection proportional
to the intensity, with the direction determined by the polarization
axis. Within the isotropic model presented here, the magnitude of
the current is insensitive to rotation of the crystal axes with respect
to the normal.

\begin{figure}
\capstart%

\includegraphics{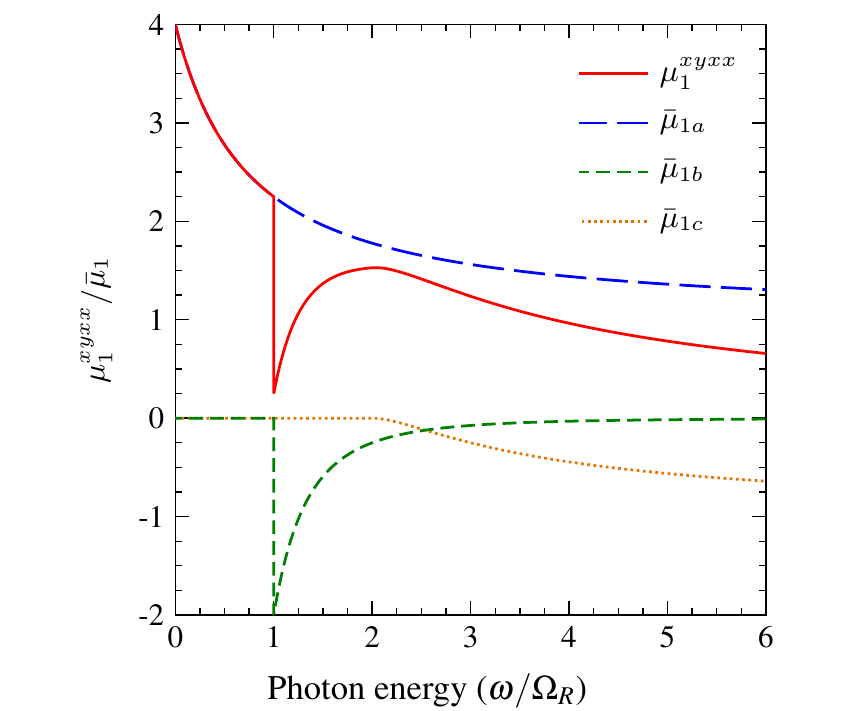}

\caption{Spin current injection strength for graphene with Rashba spin-orbit
interaction. The single independent component $\mu_{1}^{xyxx}(\omega)$
of the total spin current injection tensor {[}\prettyref{eq:mu1-graphene}{]},
as well as the individual contributions $\bar{\mu}_{1a\text{--}c}(\omega)$
{[}\prettyref{eq:mu1-components}{]}, are shown as a function of the
light frequency $\omega$ for the case of intrinsic graphene ($E_{F}=0$).
\label{fig:mu1-graphene}}
\end{figure}

The single independent component $\mu_{1}^{xyxx}(\omega)$ of the
injection tensor gives the magnitude of the resulting current. We
find that it is built up from three contributions,
\begin{multline}
\mu_{1}^{xyxx}(\omega)=\bar{\mu}_{1a}(\omega)\,\Heaviside(\omega-\omega_{F})\\
+\tfrac{1}{2}\bar{\mu}_{1b}(\omega)\,\left[\Heaviside(\omega-\omega_{F}-\Omega_{R})+\Heaviside(\omega-\omega_{F}+\Omega_{R})\right]\\
+\bar{\mu}_{1c}(\omega)\,\Heaviside(\omega-\omega_{F}),\label{eq:mu1-graphene}
\end{multline}
where $\hbar\omega_{F}\equiv2\left|E_{F}\right|$ is twice the Fermi
level and $\Heaviside(x)$ is the Heaviside step function. The three
contributions $\bar{\mu}_{1a\text{--}c}(\omega)$ correspond, in order,
to contributions arising from the following band-to-band transitions
{[}\emph{c.f.}~\prettyref{subfiga:dispersion}{]}: (a) from $v_{2}$
to $c_{1}$, (b) from $v_{2}$ to $c_{2}$ and from $v_{1}$ to $c_{1}$
with equal contribution, and (c) from $v_{1}$ to $c_{2}$. They are
given by\begin{subequations}
\label{eq:mu1-components}
\begin{align}
\bar{\mu}_{1a}(\omega) & =\bar{\mu}_{1}(\omega)\frac{\left(\omega+2\Omega_{R}\right)^{2}}{\left(\omega+\Omega_{R}\right)^{2}},\\
\bar{\mu}_{1b}(\omega) & =-2\bar{\mu}_{1}(\omega)\frac{\Omega_{R}^{3}}{\omega^{3}}\Heaviside(\omega-\Omega_{R}),\\
\bar{\mu}_{1c}(\omega) & =-\bar{\mu}_{1}(\omega)\frac{\left(\omega-2\Omega_{R}\right)^{2}}{\left(\omega-\Omega_{R}\right)^{2}}\Heaviside(\omega-2\Omega_{R}),
\end{align}
\end{subequations}
where $\bar{\mu}_{1}(\omega)\equiv e^{2}v_{F}\left(16\hbar\omega\right)^{-1}$
including the valley degeneracy.  The individual contributions $\bar{\mu}_{1a\text{--}c}(\omega)$
and the total spin current injection tensor for intrinsic graphene
are plotted as a function of the light frequency in \prettyref{fig:mu1-graphene}.

\begin{figure}
\capstart%

\includegraphics{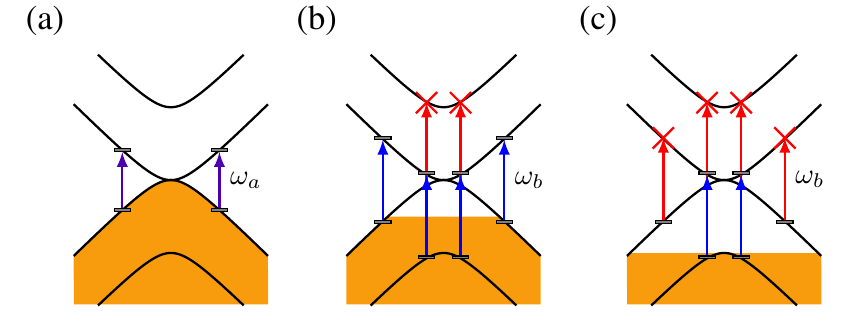}

\caption{Photoinduced pure spin current injection schemes. (a) Intrinsic graphene
at low energy, $\omega_{a}<\Omega_{R}$, yields the strongest spin
current injection. (b) For mid-range energy, $\Omega_{R}<\omega_{b}<2\Omega_{R}$,
the spin current injection strength improves by tuning the Fermi level
to $\left|E_{F}\right|=\frac{1}{2}\hbar\Omega_{R}$ to Pauli block
half of the transitions involving a split-off band. (c) For the same
frequency range, further increasing the Fermi level to $\left|E_{F}\right|=\hbar\Omega_{R}$
results in the Pauli blocking of transitions involving the gapless
bands. This yields a reversal of the spin current injection. \label{fig:injection-schemes}\label{subfiga:injection-schemes}
\label{subfigb:injection-schemes} \label{subfigc:injection-schemes}}
\end{figure}

At photon energy less than the Rashba frequency, only $\bar{\mu}_{1a}$
arising from the low-energy bands contributes to the total spin current
injection, as transitions involving other bands are energetically
forbidden. This contribution is maximal in the limit $\omega\rightarrow0$,
decreases monotonically with frequency, and mostly dictates the total
spin current injection everywhere except near the onset at $\omega=\Omega_{R}$
and in the $\omega\gg\Omega_{R}$ limit. At photon energy equal to
the Rashba frequency, $\omega=\Omega_{R}$, transitions involving
one split-off band start contributing. This contribution, $\bar{\mu}_{1b}$,
opposes the previous contribution, $\bar{\mu}_{1a}$, and has a sharp
onset where the two contributions are roughly of the same amplitude,
resulting in a sharp dip in the total spin current injection. However,
$\bar{\mu}_{1b}$ decreases to zero fairly quickly with increasing
frequency, and $\bar{\mu}_{1a}$ is again the dominant contribution.
Another onset occurs at $\omega=2\Omega_{R}$, where transitions from
the split-off valence to the split-off conduction band start occurring.
This contribution, $\bar{\mu}_{1c}$, is initially small, increases
with increasing frequency, and opposes the initial contribution, $\bar{\mu}_{1a}$.
At photon energy much larger than the Rashba frequency, $\omega\gg\Omega_{R}$,
the contributions $\bar{\mu}_{1a}$ and $\bar{\mu}_{1c}$ tend to
$\bar{\mu}_{1}$ and $-\bar{\mu}_{1}$, respectively, and the overall
spin current injection is zero.

\begin{figure}
\capstart%

\includegraphics{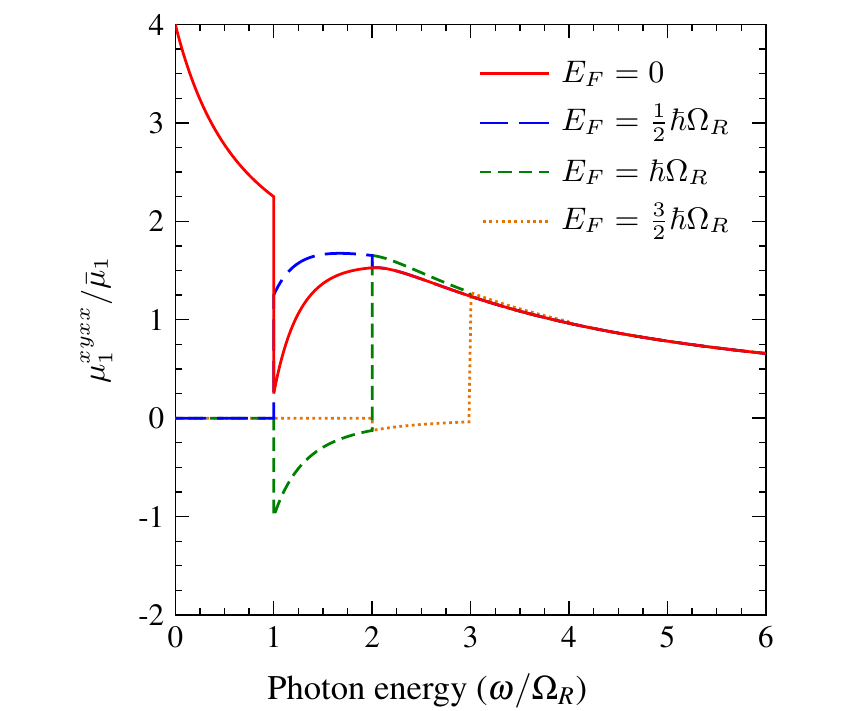}

\caption{Spin current injection strength $\mu_{1}^{xyxx}(\omega)$ for graphene
with Rashba spin-orbit interaction {[}\prettyref{eq:mu1-graphene}{]}
for varying values of the Fermi level ($E_{F}=0$, $\tfrac{1}{2}\hbar\Omega_{R}$,
$\hbar\Omega_{R}$, $\tfrac{3}{2}\hbar\Omega_{R}$). \label{fig:mu1-graphene-Fermi-level}}
\end{figure}

For intrinsic graphene, the spin current injection strength $\mu_{1}^{xyxx}(\omega)$
is positive at all frequencies. The strongest spin current injection
is achieved when $\omega<\Omega_{R}$, a situation illustrated in
\prettyref{subfiga:injection-schemes}. The spin current injection
decreases significantly for frequency above $\omega=\Omega_{R}$,
once transitions involving one split-off band start contributing.
It is possible to alleviate this effect by tuning the Fermi level
$E_{F}$ away from the charge neutrality point, as in \prettyref{subfigb:injection-schemes}.
The transition from the lowest valence band is then Pauli blocked,
and the diminishing contribution originating from $\bar{\mu}_{1b}$
is reduced. This results in an increased total spin current injection
for frequencies in the range $\omega_{F}<\omega<\omega_{F}+\Omega_{R}$.
This Pauli-blocking scheme also scales to higher energies, allowing
one to increase the photoexcitation effect in a tunable range of frequencies.
The maximal spin current injection at a given frequency $\omega$
is achieved by tuning the Fermi level such that $E_{F}=\left(2\hbar\omega-\hbar\Omega_{R}\right)/4$.
Examples of nonzero doping are given in \prettyref{fig:mu1-graphene-Fermi-level},
showing the reduction of the dip in spin current injection that occurs
for intrinsic graphene at $\omega=\Omega_{R}$ by tuning the Fermi
level to $\omega_{F}=\Omega_{R}$. For larger values of the Fermi
level, when $\omega_{F}>\Omega_{R}$, the sign of $\mu_{1}^{xyxx}(\omega)$
is also affected by the doping, taking negative values for $\omega<\omega_{F}$
and positive values for $\omega>\omega_{F}$. This occurs since the
positive and usually dominant contribution arising from the gapless
bands, $\bar{\mu}_{1a}$, is Pauli blocked for $\omega<\omega_{F}$,
as illustrated in \prettyref{subfigc:injection-schemes}. As seen
in \prettyref{fig:mu1-graphene-Fermi-level}, the benefits of the
Pauli-blocking scheme to increase the spin current injection strength
are really substantial only from $\omega=\Omega_{R}$ to about $\omega=3\Omega_{R}$.

A measure of the polarization $P$ of the injected spin current is
obtained by taking the ratio of the injection rates between spin current
injection and carrier injection, and normalizing by the maximal velocity
$v_{F}$ and spin $\hbar/2$ for the carriers. Just as the spin current
injection rate, \prettyref{eq:mu1-definition}, the injection rate
for the density of conduction electrons is proportional to the light
intensity and is written as
\begin{equation}
\dot{n}=\xi_{1}^{ab}(\omega)E^{a*}(\omega)E^{b}(\omega).\label{eq:ndot1}
\end{equation}
The response tensor $\xi_{1}^{ab}(\omega)$ is computed using Fermi's
golden rule \citep{Rioux2011a}, yielding, for the intrinsic graphene
Hamiltonian including Rashba spin-orbit coupling,
\begin{multline}
\xi_{1}^{ab}(\omega)=\frac{\sigma_{0}\delta^{ab}}{\left(\hbar\omega\right)}\Biggl(\frac{\omega+2\Omega_{R}}{\omega+\Omega_{R}}+2\frac{\Omega_{R}^{2}}{\omega^{2}}\Heaviside(\omega-\Omega_{R})\\
+\frac{\omega-2\Omega_{R}}{\omega-\Omega_{R}}\Heaviside(\omega-2\Omega_{R})\Biggr),
\end{multline}
where $\sigma_{0}$ is the universal optical conductivity of free-standing
graphene, $\sigma_{0}=e^{2}/4\hbar$ \citep{Ando2002a,Nair2008}.

\begin{figure}
\capstart%

\includegraphics{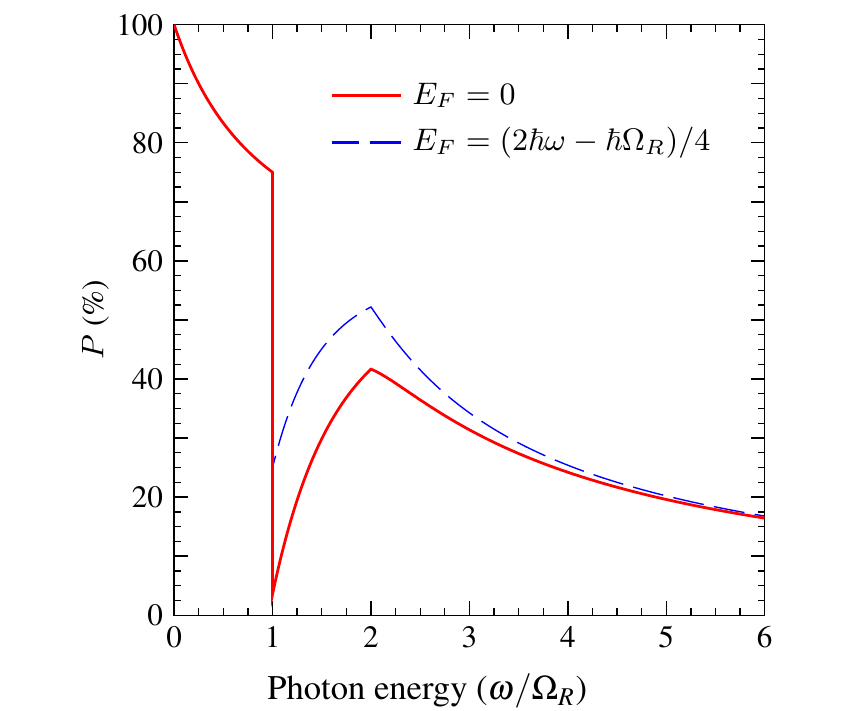}

\caption{Spin current polarization per carrier, $P$, for graphene with Rashba
spin-orbit interaction {[}\prettyref{eq:polarization}{]} for intrinsic
graphene ($E_{F}=0$) and for the maximal case making use of Pauli
blocking ($E_{F}=\left(2\hbar\omega-\hbar\Omega_{R}\right)/4$). \label{fig:mu1-graphene-DSP}}
\end{figure}

For linearly-polarized light, which yields the maximal spin current
polarization, we have
\begin{equation}
P=\frac{\left|\dot{\vec{J}}_{s}\right|}{\tfrac{\hbar}{2}v_{F}\dot{n}}=\frac{\left|\mu_{1}^{xxxy}(\omega)\right|}{\tfrac{\hbar}{2}v_{F}\left|\xi_{1}^{xx}(\omega)\right|}.\label{eq:polarization}
\end{equation}
This measure is plotted as a function of light frequency in \prettyref{fig:mu1-graphene-DSP},
showing the spin current polarization for intrinsic graphene ($E_{F}=0$)
and the maximally achievable polarization making use of the Pauli-blocking
scheme ($E_{F}=\left(2\hbar\omega-\hbar\Omega_{R}\right)/4$). Photoinduced
spin current injection with linearly-polarized light yields up to
$100$\% polarization in the limit $\omega\rightarrow0$. The polarization
decreases with increasing frequency but remains greater than $75$\%
for $\omega<\Omega_{R}$. At the onset for transitions involving a
split-off band, $\omega=\Omega_{R}$, the polarization drops sharply
to $3.6$\% ($25$\% in the maximal case) before steadily recovering
and reaching $42$\% ($52$\% in the maximal case) at the next onset,
$\omega=2\Omega_{R}$. Although the polarization then decreases monotonically,
it remains as large as $20$\% at $\omega=5\Omega_{R}$.

\section{Summary \label{sec:Summary}}

The injection of pure spin currents in graphene with Rashba spin-orbit
coupling via photoexcitation has been presented. The spin current
injection strength is zero for circularly-polarized light and maximal
for linearly-polarized light, with the spin current injection rate
given in \prettyref{eq:Jdot-linear}. The injection rate is proportional
to the light intensity and the direction of the current follows the
polarization axis. Multiple regimes of excitation have been proposed,
covering a wide range of photon frequencies. The technique achieves
very high spin current polarization, above $75$\% at frequencies
below the Rashba frequency, roughly $50$\% at twice the Rashba frequency,
and remains as high as $20$\% at five times the Rashba frequency.
In comparison, electrical injection has recently achieved $60$\%
polarization \citep{Friedman2014}.

The injection of a pure spin current is interesting for spintronics
applications, and such currents could be detected via electrical edge
currents \citep{Kane2005c}, electrical currents \citep{Vera-Marun2011,*Vera-Marun2012},
pump-probe spectroscopy \citep{Zhao2005}, or Faraday rotation \citep{Chen2010}.
Spin current injection via optical methods removes the need for ferromagnetic
contacts, which have been identified as a possible source of spin
scattering in electrical spin injection in graphene. Since carriers
are injected ballistically with high carrier velocities, on the order
of the Fermi velocity, the spin separation can reach a commensurate
distance after excitation before the spin current decays, its lifetime
limited by momentum relaxation \citep{Sherman2008}. A careful treatment
of the subsequent carrier dynamics after injection, including the
effect of disorder, inhomogeneity in doping level and SOI strength,
is of interest and the topic of future work.

The range of frequencies yielding large spin current polarization
is increased by a larger spin-orbit coupling strength. A number of
studies have reported enhanced SOI in graphene \citep{CastroNeto2009b,Marchenko2012,Balakrishnan2013,Gmitra2013}.
The optical injection of a spin current in bilayer graphene represents
another interesting avenue as it presents a stronger SOI \citep{Konschuh2012,*Konschuh2012e1,*Konschuh2012e2}.
While the dependence of the magnitude $\mu_{1}^{xyxx}(\omega)$ as
a function of light frequency for bilayer graphene is necessarily
more complicated than that of graphene, due to the additional available
interband transitions, at the simplest level an isotropic bandstructure
model can be used so that the symmetry considerations of the present
paper hold.

In previous proposals of optical spin current injection in low-dimensional
structures, spin displacement results from the interference of absorption
pathways for left and right circularly-polarized components of the
linearly-polarized light \citep{Tarasenko2005a,Sherman2005a,Bhat2005b,Zhao2005}.
Due to the completely different band symmetry, such optical orientation
of the electron spin under circularly-polarized light is not possible
in graphene without applying an additional magnetic field \citep{Inglot2014}.
Nevertheless, we have shown that optical pure spin current injection
in graphene is possible, with in-plane spin and velocity components
and without a magnetic field.

\appendix

\begin{acknowledgments}
This work was supported by the Konstanz Center for Applied Photonics
(CAP) and by the Deutsche Forschungsgemeinschaft (DFG) through SFB
767.
\end{acknowledgments}

\section*{Appendix: Matrix elements \label{secapp:Matrix-elements}}

The matrix elements necessary to calculate the spin current injection
tensor $\mu_{1}^{abcd}(\omega)$ according to \prettyref{eq:mu1-microscopic}
are as follows. The velocity operator $\vec{v}=\vec{\sigma}v_{F}$,
written in the eigenstate basis $\left\{ c_{1},v_{2},c_{2},v_{1}\right\} $,
takes the form
\begin{equation}
\vec{v}=\frac{\Omega_{R}}{\sqrt{\Omega_{R}^{2}+4v_{F}^{2}k^{2}}}\left(\begin{array}{cc}
\frac{\hbar k}{m^{*}}A & v_{F}B\\
v_{F}B^{\dagger} & \frac{\hbar k}{m^{*}}A
\end{array}\right)
\end{equation}
where
\begin{align}
A & \equiv\left(\begin{array}{cc}
\hat{k} & i\tau\hat{\phi}\\
-i\tau\hat{\phi} & -\hat{k}
\end{array}\right),\\
B & \equiv\left(\begin{array}{cc}
-i\hat{\phi} & -\tau\hat{k}\\
\tau\hat{k} & i\hat{\phi}
\end{array}\right),
\end{align}
and $m^{*}=\hbar\Omega_{R}/2v_{F}^{2}$ is the effective mass describing
the quadratic band dispersion of graphene with Rashba SOI in the limit
$k\rightarrow0$. The spin operator written in the same eigenstate
basis takes the form
\begin{equation}
\vec{S}=\frac{\hbar}{2}\frac{1}{\sqrt{\Omega_{R}^{2}+4v_{F}^{2}k^{2}}}\left(\begin{array}{cc}
-C & D\\
D^{\dagger} & C
\end{array}\right)
\end{equation}
where
\begin{align}
C & \equiv\left(\begin{array}{cc}
2v_{F}k\hat{\phi} & -\tau\Omega_{R}\hat{z}\\
-\tau\Omega_{R}\hat{z} & 2kv_{F}\hat{\phi}
\end{array}\right),\\
D\equiv & \left(\begin{array}{cc}
2v_{F}k\hat{z}+i\sqrt{\Omega_{R}^{2}+4v_{F}^{2}k^{2}}\,\hat{k} & \tau\Omega_{R}\hat{\phi}\\
\tau\Omega_{R}\hat{\phi} & 2v_{F}k\hat{z}+i\sqrt{\Omega_{R}^{2}+4v_{F}^{2}k^{2}}\,\hat{k}
\end{array}\right).
\end{align}
From these one can obtain matrix elements of the spin current operator
following \prettyref{eq:J-matrix-element}.

\def\firstpage#1--#2\relax{#1}\def\pages#1{\firstpage
  #1\relax}

\end{document}